# Modeling Three-dimensional Invasive Solid Tumor Growth in Heterogeneous Microenvironment under Chemotherapy

Hang Xie[1], Yang Jiao[2], Qihui Fan[3], Miaomiao Hai[1], Jiaen Yang[1], Zhijian Hu[1], Yue Yang[4], Jianwei Shuai[5], Guo Chen[1], Ruchuan Liu[1]\*, Liyu Liu[1]\*

**1** College of Physics, Chongqing University, Chongqing, 401331 China, **2** Materials Science and Engineering, Arizona State University, Tempe, AZ 85287, United States of America, **3** Key Laboratory of Soft Matter Physics, Institute of Physics, Chinese Academy of Science, Beijing, 100190 China, **4** Key Laboratory of Carcinogenesis and Translational Research (Ministry of Education), Department of Thoracic Surgery II, Peking University School of Oncology, Beijing Cancer Hospital and Institute, 52 Fucheng Avenue, Haidian District, Beijing 100142, China, **5** Department of Physics, Xiamen University, Xiamen 361005, China

\* phyliurc@cqu.edu.cn (RL); lyliu@cqu.edu.cn (LL)

**Abstract**

A systematic understanding of the evolution and growth dynamics of invasive solid tumors in response to different chemotherapy strategies is crucial for the development of individually optimized oncotherapy. Here, we develop a hybrid three-dimensional (3D) computational model that integrates pharmacokinetic model, continuum diffusion-reaction model and discrete cell automaton model to investigate 3D invasive solid tumor growth in heterogeneous microenvironment under chemotherapy. Specifically, we consider the effects of heterogeneous environment on drug diffusion, tumor growth, invasion and the drug-tumor interaction on individual cell level. We employ the hybrid model to investigate the evolution and growth dynamics of avascular invasive solid tumors under different chemotherapy strategies. Our simulations reproduce the well-established observation that constant dosing is generally more effective in suppressing primary tumor growth than periodic dosing, due to the resulting continuous high drug concentration. In highly heterogeneous microenvironment, the malignancy of the tumor is significantly enhanced, leading to inefficiency of chemotherapies. The effects of geometrically-confined microenvironment and non-uniform drug dosing are also investigated. Our computational model, when supplemented with sufficient clinical data, could eventually lead to the development of efficient *in silico* tools for prognosis and treatment strategy optimization.

# 1. Introduction

Cancer is a group of highly fatal diseases that usually involve abnormal cell growth and emergent migration behaviors due to complex tumor-host interactions, leading to invasion and metastasis. For a typical solid tumor, the proliferative cells take up oxygen and nutrition from surrounding microenvironment and actively produce daughter cells to expand the tumor mass. The cells in the inner region of the tumor become inactive (quiescent) due to starving and eventually turn necrotic. In malignant tumors, mutant daughter cells with invasive phenotype i.e., low cell-cell adhesion, high mobility and strong drug resistance, are produced and can detach from the primary tumor and migrate into the surrounding stromal [1-4]. Such invasive cells can enter the circulation systems (e.g., blood vessels) and reside in distant organs, leading to the emergence of secondary tumor and metastasis, and thus makes it very difficult for cancer treatment.[5]

To better understand the evolution and invasive of malignant tumors and the influence of the host microenvironment, a variety of computational models on tumor growth have been devised, which can be generally categorized continuum [6-14], discrete [15-20] and hybrid [21-26] models, to name but a few. The continuum models typically employ coupled partial differential equations (e.g., diffusion-reaction equations) characterizing tumor population evolution in homogeneous microenvironment as well as the evolution oxygen and nutrient concentrations due to cancer cell consumption and metabolism. The continuum models are able to capture the complex diffusion dynamics of the nutrients, the tumor growth and cell apoptosis as well as the effects of chemotaxis and cell adhesion, and can be easily employed to investigate large systems containing millions of cancer cells in the mature tumor. However, the detailed evolution and phenotype heterogeneity of individual tumor cells cannot be studied using the continuum models.

In the discrete models, individual tumor cells are explicitly considered and the tumor system can be represented using either the particle-assembly model[20] or the cellular automaton (CA) model[15-19]. In the particle-assembly model, each tumor cell is represented as a bag of incompressible fluid enclosed by a hyper-elastic membrane with prescribed properties, which can capture detailed morphology evolution of the entire proliferative colony. In the CA model, the simulation domain is pre-tessellated into "automaton cells", and each automaton cell is assigned a value representing either a biological cell in a particular state (e.g., proliferative, quiescent or necrotic) or a region of host microenvironment. The state of a specific automaton cell depends on those of the neighboring cells via prescribed CA rules. The original CA models were devised to simulate the proliferative growth of brain tumors [16] and have been recently generalized to investigate phenotype heterogeneity, invasive growth [27,28], effects of confined heterogeneous environment [15,28], angiogenesis [17,22], and tumor dormancy [29]. The hybrid models typically integrate the continuum model for nutrient concentration evolution and the CA model for individual cell dynamics, explicitly considering the coupling of the two via nutrients up-take and consumption for cell proliferation[21-24]. Due to the computational cost, most existing hybrid models are focused on 2D systems. The readers are referred to recent reviews for a more detailed discussion of the aforementioned tumor simulation models.[30-33]

An outstanding issue in oncotherapy is the lacking of a systematic understanding of the evolution and growth dynamics of invasive solid tumors in response to different chemotherapy strategies. Such an understanding is crucial for the development of individually optimized oncotherapy. Typical chemotherapeutic agents (drugs) interfere with cancer cell division (mitosis) to cause cell damage or death, suppressing the overall growth of the tumor[34,35]. Generally, drug macromolecules are transported to the tumor site via diffusion in the stromal and then up-taken by the tumor cells. The effectiveness of chemotherapy strongly depends on the drug concentration around the tumor cells. However, a high drug concentration also damages normal and healthy tissue cells, leading to significant side effects for the patient. An optimized dosing strategy that can result in efficient elimination of tumor cells while maintaining the integrity and functionality of normal healthy tissues is crucial to the success of the cancer treatment. Periodic dosing has recently been suggested as a promising treatment strategy [36]. In order to maximize the treatment effectiveness, the heterogeneity of the tumor-host system[37,38] as well as the variation in the drug's cytotoxicity (cell killing) and the effects of tumor hypoxia [34,35] should also be taken into account.

In most chemotherapy, the anti-tumor drugs are either absorbed in the digestion system or directly injected into the circulation systems, and then transported different organs and tumor sites by the blood vessels. The drugs then diffuse into the avascular tissues while being metabolized by cells. The evolution of average drug concentrations in plasma, interstitial tissues and different organs can be captured via the pharmacokinetic (PK) calculations[39-41]. Such PK calculations involve master ordinary different equations (ODE) that take into the consumption of drugs due to decomposition and metabolism, as well as transport of drugs between different counterparts (e.g., organs) in the body. Although the PK models can provide average drug concentration in different organs, it is not able to describe the detail temporal-spatial evolution of drug concentration within an organ or tissue. To solve such temporal-spatial evolution, diffusion-reaction models are usually employed, in which the consumption of drugs is quantified via the "sink" terms in the associated partial differential equations (PDE).

Recently, significant research efforts have been devoted to computational modeling of various aspects of chemotherapy. For example, the effects of spatial heterogeneity in drug concentration[42], vascular structure and heterogeneous host environment[7,43], cell packing density[44], intrinsic heterogeneity in cell phenotypes and cell cycles[45-50] on the effectiveness of treatment and acquired drug resistance[51,52] have been systematically studied. Computational tools for treatment optimization have been devised[53-56] and data-based platform has been developed to assess robustness of treatment[57].

In this work, we develop a hybrid three-dimensional computational model that integrates the physiologically based pharmacokinetic model, continuum diffusion-reaction model and discrete cell automaton model to investigate 3D invasive solid tumor growth in heterogeneous microenvironment under chemotherapy. In particular, we explicitly consider the effect of heterogeneous environment on drug diffusion, tumor growth and invasion as well as the drug-tumor interaction on individual cell level. We employ the hybrid model to investigate the effectiveness of two commonly used dosing strategies, i.e., constant and periodic dosing, in controlling the growth of avascular invasive solid tumors. Our model robustly reproduces the observation that constant dosing is generally more effective in suppressing primary tumor growth compared to periodic dosing, due to the resulting continuous high drug concentration[58-61]. However, the suppression of primary tumor progression does not necessarily lead to a suppression of invasive cell migration, which results in complex invasion branches emitting from the primary tumor[62-65]. Moreover, we show that microenvironment heterogeneity can significantly enhanced the malignancy of the tumor and thus, reduce the effectiveness of the chemotherapy with even periodic dosing[66-69].

## 2. Materials and Methods

In our model, the computational domain is a sub-region in an organ that contains both the growing avascular tumor with possible invasive branches and the surrounding stroma. Specifically, we consider two types of the host micro-environment respectively with avascular tissues (Fig. 1(a)) and with vascular tissues (Fig. 1(b)). We consider that within the simulation domain, there are no blood vessels and the drugs are transported to the tumor region via diffusion. The drug contraction in the organ will first be obtained via the physiologically based pharmacokinetic (PBPK) model and imposed as boundary condition at the boarder of the computational domain.

The evolution of drug concentration within the computational domain is described by a diffusion-reaction equation, which includes position dependent diffusion coefficient for heterogeneous microenvironment and consumption terms characterizing the drug's metabolism and decomposition. Different dosing strategies are simulated using different time-dependent boundary conditions. The evolution of the invasive solid tumor is simulated using the cellular automaton model. Specifically, the probability of division of each proliferative cell is now considered a function of not only the local microenvironment (e.g., ECM density, rigidity and pressure) but also the local drug concentration computed using the diffusion-reaction equation. We consider the migration an invasive cell depends only on the microenvironment. In the subsequent subsections, we will provide detailed descriptions of these models and their integration.

## 2.1 Pharmacokinetic modeling of temporal-evolution of overall drug concentration

In most chemotherapy, the anti-tumor drugs are either absorbed in the digestion system or directly injected into the circulation systems, and then transported to different organs and tumor sites by the blood vessels. The decay of average drug concentrations over time in different organs (due to transport of drugs among different organs and drug consumption) is usually described by a set of coupled ordinary differential equations referred to as the physiologically based pharmacokinetic (PBPK) models [39,41,70]. Here, we apply a two-compartment PBPK model, in which the drug concentrations in a vascularized compartment ($C_1$) and the concentration in an avascular compartment ($C_2$) are considered. For the vascularized compartment, e.g., a tumor micro-environment with a high density of blood vessels in the stromal tissue, the transport of drugs is mainly through blood vessel. For the avascular compartment, e.g., an avascular tumor, the drug transportation is mainly via diffusion in the stromal. The corresponding PBPK equations characterization the temporal evolution of $C_1$ and $C_2$ are given below:

$$\frac{dC_1}{dt} = k_{21}\frac{V_2}{V_1}C_2(t) - (k_{12}+k_{10})C_1(t) + \frac{d(t)}{V_1} \qquad (1)$$

$$\frac{dC_2(t)}{dt} = k_{12}\frac{V_1}{V_2}C_1(t) - k_{21}C_2(t) \qquad (2)$$

where $k_{21}$, $k_{12}$, $k_{10}$ are transport rate constants (for example, $k_{21}$ is the transport rate from compartment 2 to compartment 1); $V_1$ and $V_2$ are volumes of the compartments, $d(t)$ is the time-dependent dosage as a drug source. Eq. (1) means that the change rate of $C_1$ consists of the incoming flow from the compartment 2 (the first term); the outgoing flow (the second term) and the injection flow from the dosage source (the last term).

Using the same set of model parameters and initial conditions given in Ref. [39] (see the values in the caption of Fig. 1(c)), we can calculate of the average drug (CPT-11) concentration as a function of time in the vascularized and avascular compartments, representing respectively a vascularized and avascular tumor environment, for a given dosing condition. Fig. 1(c) shows the results for an impulse dosing at t = 0. It can be seen that in both compartments, the average drug concentration decays monotonically with time and $C_1$ in the vascularized compartment possessing a much higher initial value, also drops much faster than $C_2$ in the avascular compartment. This is due to the different drug transport mechanism. These PBPK results are imply that the drug concentration in tumors in avascular micro-environments decays much slower than that in those in vascularized micro-environments, although the initial concentration value in the avascular micro-environment is lower than that in the vascularized micro-environment for the same initial dosage.

Based on the PBPK calculations, in our subsequent simulations, two distinct types of time-dependent boundary conditions characterizing the drug concentration at the boundary of our simulation domain will be used respectively for tumors in vascularized and avascular micro-environments. In particular, we consider that for the avascular micro-environment, the drug concentration has very slow decay after each bolus injection; and in the vascularized micro-environment, the drug concentration decays quickly after a bolus injection. On the other hand, for constant dosing, the drug is continuously supplied leading to an almost constant concentration level of the drugs in different compartments.

**2.2 Diffusion-reaction model for temporal-spatial evolution of drug concentration in tumor systems**

To simulate the temporal-spatial evolution of drug concentration in tumor systems, we employ the following diffusion-reaction model:

$$\frac{\partial \phi}{\partial t} = D_0 \nabla^2 \phi - K_{met}\phi - \lambda_0 \frac{\phi}{\phi + \phi_0} \tag{3}$$

where $D_0$ is the diffusion coefficient of the drug; the last two terms result from drug consumption. Specifically, the parameter $K_{met}$ is the first-order decay rate due to the chemical decomposition as the drug macromolecules diffuse in the stroma. The last term on the right hand side of Eq. (3), usually referred to as the Michaelis-Menten metabolism term [23,40], characterizes the drug up-take by tumor cells, i.e., drug concentration will decrease in the presence of the tumor cells, and we set $\lambda_0 = \lambda_{14}n$, where $n$ is the tumor cell number density, which is computed from CA model as described below.

We consider the simulation domain is initially drug free and the drugs enter the domain through the boundary. For constant dosing, a time-independent drug concentration value will be used of the boundary condition. For periodic dosing, a general time-dependent boundary condition suggested by the PBPK calculations is employed to capture the time-evolution of drug concentration in the tumorous organ due to pharmacokinetics as well as different dosing cycles. Specifically, in the early stage of one dosing cycle, i.e., the drug infusion period characterized by the infusion time $\tau_{infusion}$, the drug concentration approximately remains a level in the system and then decays to zero after $\tau_{infusion}$ as indicated by pharmacokinetics. This infusion-decay process corresponds to a complete dosing cycle with a period $\tau_{cycle}$ and such a process is repeated to simulate periodic dosing. For different tumor micro-environment (i.e., vascularized vs. avascular), the rate of decay for the drug concentration is taken differently according the PBPK calculations.

To numerically solve the diffusion-reaction equation (3), we use the finite difference method. In particular, the simulation domain containing the solid tumor and stroma is discretized into a cubic grid with $N_f^3$ points. The Euler forward-finite difference scheme is used, i.e.,

$$\phi_{i,j,k}^{n+1} = \phi_{i,j,k}^n + \Delta t \cdot [D_0 \Delta(\phi_{i,j,k}^n) - K_{met}\phi_{i,j,k}^n - \lambda_{14} n_{i,j,k} \frac{\phi_{i,j,k}^n}{\phi_{i,j,k}^n + \phi_0}] \quad (4)$$

where the Laplacian operator for spatial finite difference is written as

$$\Delta(\phi_{i,j,k}^n) = (\phi_{i+1,j,k}^n + \phi_{i-1,j,k}^n + \phi_{i,j+1,k}^n + \phi_{i,j-1,k}^n + \phi_{i,j,k+1}^n + \phi_{i,j,k-1}^n - 6\phi_{i,j,k}^n)/\Delta x^2. \quad (5)$$

This algorithm is numerically stable for the pure diffusion equation, if the following condition is satisfied [71]

$$\frac{D_0 \Delta t}{\Delta x^2} < \frac{1}{2} \quad (6)$$

We have verified that Eq. (6) is also sufficient to guarantee numerically stability for Eq. (4). $n_{i,j,k}$ is the cell density from the CA model, as is described later.

The obtained drug concentration value on each grid point is then mapped to the nearest automaton cells, which is then used to determine the decrease factor for the division rate of the proliferative cells in the CA model. The number of automaton cells representing the tumor cells within the volume element associated with a grid point is also obtained and used to calculation the tumor cell number density *n* for Eq. (3). The detailed implementation for coupling the continuum diffusion-reaction model and the discrete CA model is provided in Sec. 2.4.

**2.3 Cellular automaton model for invasive tumor growth in heterogeneous microenvironment under effects of drugs**

We employ the cellular automaton (CA) model to simulate the evolution of invasive tumor in heterogeneous microenvironment under the effects of chemotherapy. Our CA algorithm follows closely that described in Refs. [27] and [28]. In particular, the simulation domain is tessellated into polyhedra (or polygons in 2D) associated with a prescribed point configuration (i.e., the centers of randomly packed congruent hard spheres) via Voronoi tessellation. Each polyhedron is defined as an automaton cell, which in our model can either represent a real biological cell or a region of tumor stroma (which consists of a cluster of ECM macromolecules). The tumor cell can be proliferative, quiescent, necrotic or invasive in our model (see details below). Accordingly, the tumor-associated automaton cell can take distinct numerical values representing the different tumor cell state. Each ECM-associated automaton cell possesses a local density value $\rho_{ECM}$ to take into account the ECM heterogeneity, which is also positively correlated with the local ECM rigidity. When an ECM-associated automaton cell is taken by a tumor cell due to either proliferative growth or invasion, we set its $\rho_{ECM} = 0$.

In our simulation, the tumor cells in the proliferative rim (mainly in the outer shell of the primary tumor which has access to the nutrients such as oxygen and glucose) can produce daughter cells taking nearby ECM-associated automaton cells via cell division. This process leads to the growth and expansion of the primary tumor mass. The tumor cells in the inner region may turn into quiescent (alive but inactive) and then necrotic (dead) if they could not acquire sufficient nutrients for a long time. A small fraction of daughter cells may acquire invasive phenotype (e.g., weak cell-cell adhesion, strong mobility and ECM degradation ability) via mutation, which can leave the main tumor body and immigrate into the surrounding tissue leading tumor invasion. In our simulation, time is discretized into days, and for each day, the state of all tumor cells are checked for possible update. During each day, the evolution of the tumor is simulated by applying the following cellular automaton (CA) rules:

1. The quiescent cells more than a certain distance $\delta_n$ from the tumor surface are turned necrotic due to starving. The critical value of $\delta_n$ is given as

$$\delta_n = aL_t^{(d-1)/d} \tag{7}$$

where $a$ is a prescribed scaling parameter (see Table 1) and $d$ is the spatial dimension. $L_t$ is the distance between the geometric centroid ($\mathbf{x}_c$) of the tumor and the tumor edge cell that is closest to the quiescent cell considered. $\mathbf{x}_c$ is defined as $\mathbf{x}_c = \frac{1}{N}\sum_{i=1}^{N}\mathbf{x}_i$, where $N$ is the total number of noninvasive tumor cells, which is updated when a new noninvasive daughter cell is added to the tumor.

2. Each proliferative cell can produce a daughter cell the probability $P_{div}$, which will occupy an ECM-associated automaton cell in the surrounding stroma. We consider that the probability of division depends on both the heterogeneous environment and local drug concentration and possess the following expression [27,28]

$$P_{div} = p_0 \cdot P_{\gamma,\varphi} \cdot \frac{(1-\frac{r}{L_{max}})+(1-\rho_{ECM})}{2} \tag{8}$$

where $p_0$ is the base probability of division, $P_{\gamma,\varphi}$ is the cellular division reduction factor due to the chemotherapy and is considered a linear function of normalized local drug concentration ($\varphi$), i.e., $P_{\gamma,\varphi} = 1 - (1- P_{\gamma}) \varphi$; $r$ is the distance of the dividing cell from the tumor centroid; $L_{max}$ is the distance between the tumor centroid and the closest growth-permitting boundary cell in the tumor growth direction. Eq. (8) implies that $P_{div}$ depends on both the physical confinement imposed by the boundary of the growth-permitting region and the local mechanical interaction between tumor cell and the ECM, as well as the local drug concentration.

3. A proliferative cell turns quiescent if it is more than a certain distance $\delta_p$ from the tumor surface or there is no space for the placement of the forthcoming daughter cells. The distance $\delta_p$, which corresponds to the thickness of nutrient-rich proliferative rim of the primary tumor, is given by

$$\delta_p = bL_t^{(d-1)/d} \tag{9}$$

where $b$ is a prescribed scaling parameter (see Table 1), $L_t$ is the distance between the tumor centroid and the tumor edge cell that is closest to the proliferative cell considered.

4. A newly produced daughter cell can gain invasive phenotype (weak cell-cell adhesion, high motility and strong ECM degradation ability) and become an invasive cell with the mutant probability $\gamma$ (see Table 1).

5. A mutant invasive cell has the ability to degrade the nearby ECM and migrate into the surrounding stroma. We consider that the invasive cell has the mobility: $\mu$, which is the upper bound on the number of attempts the cell makes to degrade the surrounding ECM and migrate into the ECM-associated automaton cell. For example, an arbitrary invasive cell can make $m$ attempts to degrade ECM and move, where $m$ is an arbitrary integer in $[0, \mu]$. In each degradation/moving attempt, the density of the ECM-associated automaton cell will be decreased by $\delta_p$, where $\delta_p$ is an arbitrary real number in $[0, \chi]$ characterizing the cell's ECM degradation ability. After $m$ attempts, if the ECM in the automaton cell is completely degraded ($\rho_{ECM} \leq 0$), the invasive cell will migrate into this automation cell, leaving behind a path composed of degraded ECM-associated cells. The direction of motion is the one that maximizes the nutrient concentration.

6. A migrating invasive cell does not divide anymore.

| Symbols | Definition | Expression or values |
|---|---|---|
| $L_t$ | Local tumor radius (varies with cell positions) | See the text |
| $L_{max}$ | Local maximum tumor extent (varies with cell positions) | See the text |
| $a$ | Base necrotic thickness | 0.12 |
| $b$ | Base proliferative thickness | 0.08 |
| $p_0$ | Base probability of division | 0.192 |
| $\delta_n$ | Characteristic living-cell (necrotic cell) rim thickness | See Eq. (7) |
| $\delta_p$ | Characteristic proliferative rim thickness | See Eq. (9) |
| $P_{div}$ | Probability of division for proliferative cells | See Eq. (8) |
| $\gamma$ | Mutation rate for invasive cells | 0.05 |
| $\chi$ | ECM degradation ability | 0.15 |
| $\mu$ | Mobility of invasive cells | 3 |

**Table 1. Summary of the definition and numerical values of the parameters for the CA model.**

**2.4 Spatial-temporal coupling of the diffusion-reaction and cellular automaton models**

In order to investigate the effects of chemotherapy on tumor growth, one needs to couple the diffusion-reaction model describing the drug diffusion and consumption with the CA mode for tumor dynamics. Although the spatial coupling of the two models is straightforward by mapping the computational grid points for PDEs to the CA cell positions, the temporal coupling can be nontrivial.

We first estimate the characteristic drug diffusion time in the tumor. From the analytic solution of diffusion equation in a homogenous medium: $\phi(\vec{r},t) = \frac{1}{\sqrt{(4\pi Dt)^3}} e^{-\frac{r^2}{4Dt}}$ for the initial condition $\phi(\vec{r},0) = \delta(r)$ [72], we can define a characteristic "diffusion time" $t_m = \frac{x_0^2}{4D}$, at which the concentration at $x_0$ is high enough (about $e^{-1} \approx 0.37$ compared to that at the source). If we set $D = 1.3 \times 10^{-6}$ cm$^2$s$^{-1}$, $x_0 = 0.3$ cm, we see $t_m = 1.73 \times 10^4$ s, or about 0.2 day. If we set $D = 1.3 \times 10^{-7}$ cm$^2$·s$^{-1}$, $x_0 = 0.3$ cm, we have $t_m = 1.73 \times 10^5$ s or about 2 days. Comparing the common cell cycle time (ranging about 8 hours to 24 hours), we see that for the ordinary drug, the characteristic diffusion time is typically less than one cell cycle.

The above analysis suggests that the drug can diffuse rapidly into the tumor, and thus the drug concentration may change significantly in one cell cycle due to cellular uptake of the drug macromolecules and metabolism. However, the traditional CA model usually sets the cell division rate ($P_{div}$) as a constant in one cell cycle. So in our hybrid model, it is not suitable to do the drug diffusion calculation between two cell cycles. In the actual system, drug diffusion and cell division occur simultaneously and can have instant effects on one another: the drug reduces the cell division rates and the growing tumor result in drug consumption and variation of the diffusion coefficient.

In our model, we develop a quasi-parallel algorithm for coupling these two processes: We divide one cycle of the proliferation process (during which cell divides) and one dosing period (during which the drug diffuses into the tumor mass) into the same number of ($N_p$) steps. In each step, these two dynamical processes are simulated in sequence for different iterations. Then at the end of each step, the updated drug concentration distribution obtained from the diffusion-reaction model is passed to the CA model to compute updated cell division reduction factor; and the updated cell density distribution and heterogeneous diffusion coefficient distribution obtained from the CA model are passed to the diffusion-reaction model. Figure 1 (d) schematically illustrates this procedure. We note that the aforementioned procedure not only enables easy parallelization of our hybrid algorithm but also more realistically mimics the actual tumor proliferation process compared to traditional CA method. In particular, in traditional CA method, the fact that tumor cells divide at different times within a proliferation cycle is not explicitly considered. Here, by coupling a sub-spatial region of the tumor with a sub-temporal process of drug diffusion, we consider that the cells within this sub-region divide during the time span in which the diffusion occurs. This implies that tumor cells in different sub-region divide at different times. When $N_p$ is sufficiently large (e.g., = 50 ~ 100), the quasi-parallel algorithm can approximate the actual coupled processes. At last, we notice that in this coupling algorithm, we calculate the cell density in CA model by the expression: $n_{i,j,k} = \frac{1}{\Delta V}\sum_m n_m$, where $n_m$=0 or 1, for all the tumor cells in the neighbor of a finite difference grid point $(i,j,k)$, $\Delta V$ is the unit volume of the grid. This is used in the consumption calculation for Eq. (4). We use a mapping scheme to relate these quantities (such as $n_{i,j,k}$, $\phi_{i,j,k}^n$, $\rho_{ECM,(i,j,k)}^n$) between the finite difference grid and the Voronoi tessellation.

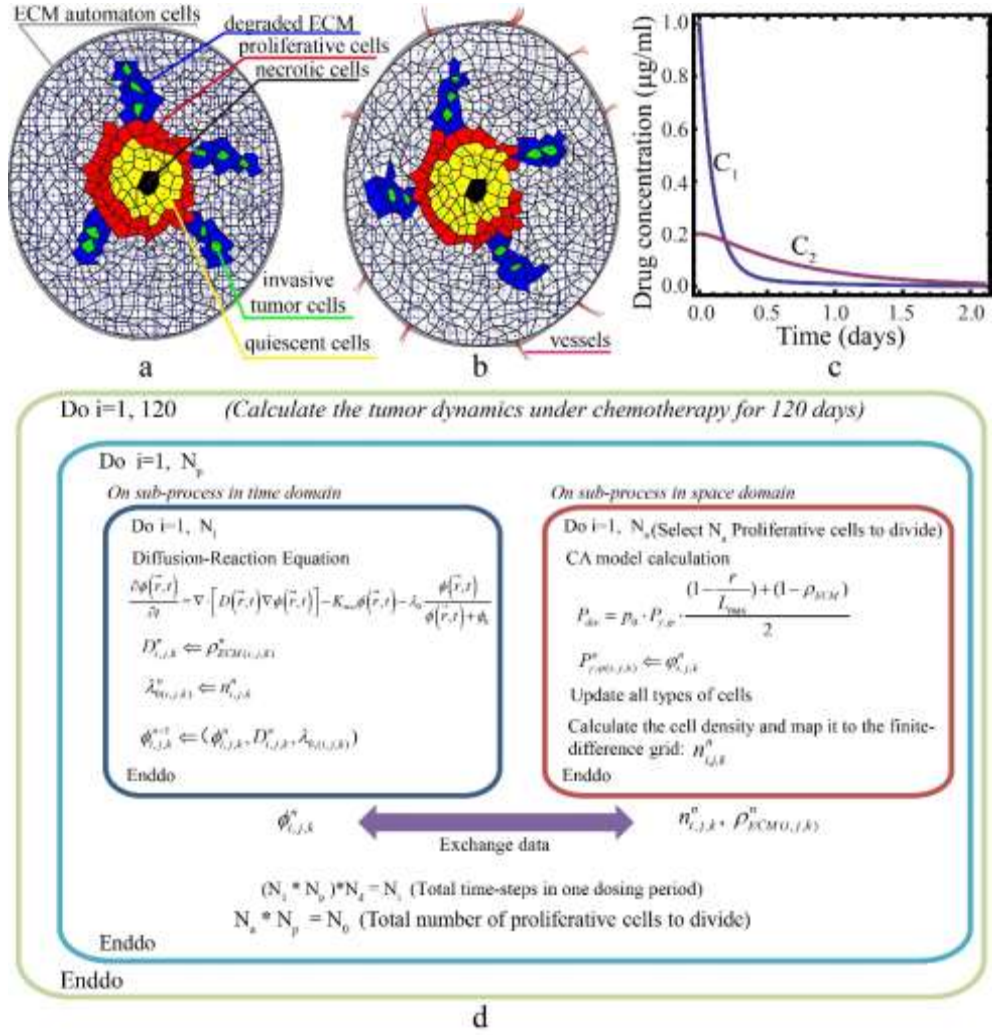

**Fig. 1 Schemes of the tumor-drug model and the time-dependency of drug concentration.** (a) and (b): 2D schematic illustrations of the tumor-drug models in this paper for the avascular tumor in an avascular (a) and vascular (b) host micro-environment, respectively. The Voronoi polyhedral in the figure is used in the CA model. The necrotic cells are black, quiescent cells are yellow, proliferative cells are red and the invasive tumor cells are green. The ECM associated automaton cells are white and the degraded ECM is blue. And the overlapped square grid (blue) is used in the finite difference calculation for the drug diffusion. (c) The drug (CPT-11) concentration dependence on time in the plasma ($C_1$) and the avascular tumor ($C_2$) calculated by the two-compartmental PBPK model (Eq. (1)-(2)). The parameters used in the calculation are: $k_{21} = 1.48$ day$^{-1}$; $k_{12} = 0.276$ day$^{-1}$; $k_{10} = 13.27$ day$^{-1}$; $V_1 = 4.85 \times 10^3$ ml; $V_2 = 8.0 \times 10^3$ ml. The initial condition at $t = 0$ is set as $C_1 = 1.0$ μg/ml and $C_2 = 0.2$ μg/ml. (d) Schematic illustration of the hybrid-parallel algorithm coupling the drug diffusion/consumption and cell division processes. The Fortran commands (do $i = 1, N$; enddo) in the rectangles indicate the loops in our algorithm.

## 3. Results and Discussions

In this section, we first study the drug dynamics in a steady-state tumor (i.e., with constant cell density distribution), in order to understand the spatial-temporal evolution of drug concentration within one proliferation cycle for both constant dosing and periodic dosing conditions. Then we employ the model to investigate the growth of avascular tumors in both vascularized and avascular homogeneous environments under constant and period dosing conditions. Finally, the coupled hybrid model is employed to study the effects of periodic dosing conditions on invasive tumor growth in heterogeneous microenvironment.

**3.1 Spatial-temporal evolution of drug concentration in steady-state tumors**

**3.1.1 Drug dynamics in steady-state tumor with constant dosing condition**

In the constant dosing condition, the boundary condition is set as a time-independent constant in the outer spherical shell (with the radius $R = 0.4$ cm) without any decay due to pharmacokinetics and is zero within the tumor region. In addition, we assume that stead-state tumor possess an ideal isotropic morphology, and tumor cell density distribution as one moves away from the tumor center can be characterized by a Fermi function ($1/(1 + \exp[(r - r_0)/\sigma])$, where $r_0$ is the tumor radius and $\sigma$ is the effective boundary-layer thickness.

The diffusion-reaction equation (3) is employed to obtain the drug concentration distribution as a function time. In particular, we consider tumors with different sizes $r_0$ and drug consumption ratio $\lambda_{14}$. Fig. 2(a) and (b) show respectively the distribution of drug concentration in a steady-state tumor with $r_0 = 0.2$ cm and $\sigma = 0.01$ cm at $t = 100$ minutes and $t = 5$ hours for different consumption ratios ($\lambda_{14} = 2.5 \times 10^{-4}$ s$^{-1}$, $\lambda_{14} = 2.5 \times 10^{-5}$ s$^{-1}$ and $\lambda_{14} = 0$). When there is no consumption, the evolution of the drug concentration is entirely controlled by the diffusion and chemical decomposition ($K_{met}$) terms, and the concentration values in the inner tumor region is larger compared to the other two cases. In addition, it is clear that a larger consumption ratio leads to a lower drug concentration in the inner tumor region. The long-time drug concentration shown in Fig. 2(b) represents steady-state solution to Eq. (3). We note that the typical time to achieve this steady-state (5 hours in the current case) is faster than a typical cell proliferation cycle, which is consistent with our time-scale analysis discussed in Sec. 2.4. Fig. 2(c) shows the drug concentration distribution in a smaller tumor ($r_0 = 0.06$ cm and $\sigma = 0.01$ cm) as a function of time with a large drug consumption ratio $\lambda_{14} = 2.5 \times 10^{-4}$ s$^{-1}$. Compared to Fig. 2(b), the steady-state of drug concentration distribution is established in a longer period of time, i.e., at $t = 1000$ minutes. This suggests that drugs need to diffuse through a wider region to achieve a steady distribution. Due to fast diffusion of drug and rapid establishment of the steady-state of drug concentration in the tumor region, we may approximate the actual drug distribution as a uniform constant for the constant dosing condition, as shown in later Sec. 3.2.1.

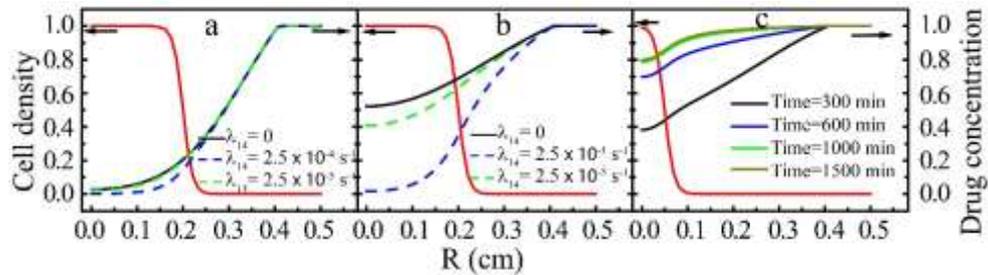

**Fig. 2 Spatial-temporal evolution of the drug concentration in the tumor model.** (a) and (b) the drug concentration distribution in an ideal tumor with $r_0 = 0.2$ cm and $\sigma = 0.01$ cm. Under constant dosing condition with different drug consumption ratios, the concentration distribution is plotted at $t = 100$ minutes (a) and 5 hours (b), respectively; (c) the drug concentration distribution in an ideal tumor with $r_0 = 0.06$ cm and $\sigma = 0.01$ cm as a function of time with a large drug consumption ratio $\lambda_{14} = 2.5 \times 10^{-4}$ s$^{-1}$ under constant dosing condition. The model parameters are set as follows: diffusion coefficient $D_0 = 1.3 \times 10^{-6}$ cm$^2 \cdot$s$^{-1}$; chemical decomposition ratio $K_{met} = 2.0 \times 10^{-4}$ min$^{-1}$. $\Delta t = 0.1$ s; $\Delta x \approx 0.01$ cm. The red lines are the corresponding cell density.

### 3.1.2 Drug dynamics in steady-state tumors with periodic dosing condition

We now consider the periodic dosing condition and employ a time-dependent periodic boundary condition to simulate the dosing condition [36]. For the periodic dosing, due to different pharmacokinetics in vascularized and avascular host micro-environments (see discussion in Sec. 2.1), we will consider these two cases separately. In particular, for the vascularized micro-environment, the drug is transported to the tumor region by blood vessels. For the avascular micro-environment, the drug reaches the tumor region mainly via diffusion. In the former case the drug concentration decays very quickly with an initial high very; and in the latter case, the drug concentration decays relatively slowly. In our simulation, we model these two cases by using different decay time parameters in the time-dependent boundary condition. Specifically, we assume that the drug concentration has the following time dependency

$$C(t) = \frac{C_0}{1+\exp[(t-\tau_{decay})/(\tau_{decay}/10)]}, \qquad t \in [N\tau_{cycle}, (N+1)\tau_{cycle}] \qquad (10)$$

where $\tau_{cycle}$ is the dosing period and $\tau_{decay}$ is the decay time for the drug in the tumor (for taking into account the pharmacokinetics effects).

Fig. 3 shows the spatial-temporal evolution of drug concentration in ideal isotropic tumors with $r_0 = 0.2$ cm and $\sigma = 0.01$ cm for both vascularized [(a) and (b)] and avascular [(c) and (d)] micro-environments under different periodic dosing conditions. Specifically, two periodic dosing conditions are applied, which are shown as the black curves in Fig. 3. We clearly can see that in the vascularized micro-environment, the drug decays rapidly ($\tau_{decay}$ = 600 mins) and in avascular micro-environment the drug decays slowly ($\tau_{decay}$ = 1800 mins). Moreover, in the avascular case, the drug is difficult to escape due to the pharmacokinetic analysis as shown in the black curves (on the tumor boundaries), the average drug concentration within the tumor always maintains in a relatively high level. This gives a positive effect on chemotherapy. Due to the decomposition and the small $\tau_{decay}$ (600 mins) leads to a very rapid decay of the drug concentration in the tumor, consistent with the pharmacokinetic analysis (see Fig. 1b). In an avascular micro-environment, the large decay time ($\tau_{decay}$

=1800 mins) leads to a slow decay of the drug within the tumor. Moreover, in the avascular case, the average drug concentration within the tumor always maintains in a high level. Due to the diffusion and consumption, the drug concentrations in the central tumor region exhibit smaller values as shown by the red dashed curves in Fig.3. With smaller drug consumption, the maximum concentration in the tumor center is larger, which is consistent with the cases for constant drug dosing conditions. Finally, we observe that there is a phase shift for the periodic variation of dosage (black curves) and drug concentration in the tumor (red curves). This is because that drug needs some time to diffuse from the outer boundary to the inner region.

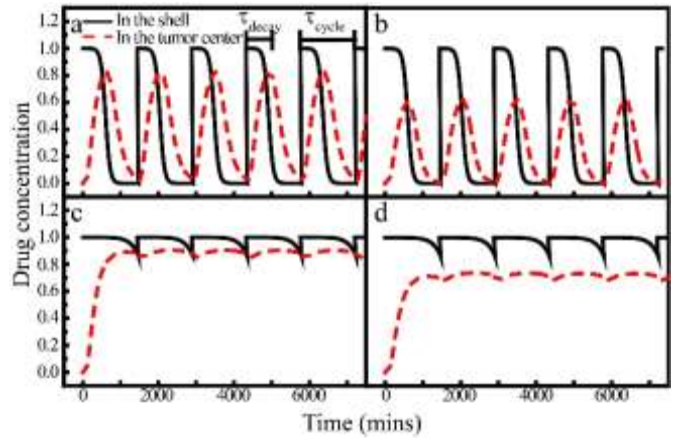

**Fig. 3 Spatial-temporal evolution of the drug concentration in ideal isotropic tumors under different periodic dosing conditions.** (a) and (b) are associated with the vascularized micro-environment with quick drag decay ($\tau_{decay}$ = 600 min); (c) and (d) are associated with the avascular micro-environment with a slow drag decay ($\tau_{decay}$ = 1800 min). (a) and (c) are associated with a small drug consumption ratio ($\lambda_{14}$ = 2.5 × 10$^{-5}$ s$^{-1}$); (b) and (d) are associated with a large drug consumption ratio ($\lambda_{14}$ = 2.5 × 10$^{-4}$ s$^{-1}$). The chemical decomposition parameter is the same in all cases: $K_{met}$ = 2.5 × 10$^{-4}$ min$^{-1}$. The black curve indicates the applied time-dependent boundary condition and the red curve indicates the drug concentration in the tumor center region. The dosing period is set as $\tau_{decay}$ = 1 day = 1440 min. The tumor size parameters are set as $r_0$ = 0.2 cm and $\sigma$ = 0.01 cm.

### 3.2 Evolution of invasive tumors in homogeneous microenvironment under chemotherapy

#### 3.2.1 Invasive tumor growth under constant dosing condition

We now employ the hybrid model to study the growth dynamics of invasive tumor under constant dosing condition in homogeneous microenvironment. To simulate the constant dosing condition, we apply a time-independent constant drug concentration at the boundary of the simulation domain. As discussed in Sec. 3.1.1, in this case, the drug concentration evolution is the same for both vascularized and avascular micro-environments as the pharmacokinetics does not play a role here. The effects of different drug concentrations are taking into account by using different cell division reduction factor $P_\gamma^0$ (= 0.6 and 0.3). Here $P_{\gamma,\varphi}$ in the CA rule 2 in Sec. 2.3 is spatial independent, $P_{\gamma,\varphi} = P_\gamma^0$). We chose the cell cycle time as one day. In the subsequent simulations, we focus on the early growth stages of the tumor.

The growth dynamics of proliferative tumors, i.e., the tumor size (radius) as a function of time is shown in Fig. 4(a). We can see that when the drug is infused in the tumor, its growth is significantly suppressed. Higher drug concentration (i.e., associated with a larger of division reduction $P_\gamma^0$) leads to the slower growth of the tumor. This is expected as the drug can significantly suppress the division of proliferative cells which is a determinant factor for the tumor growth process.

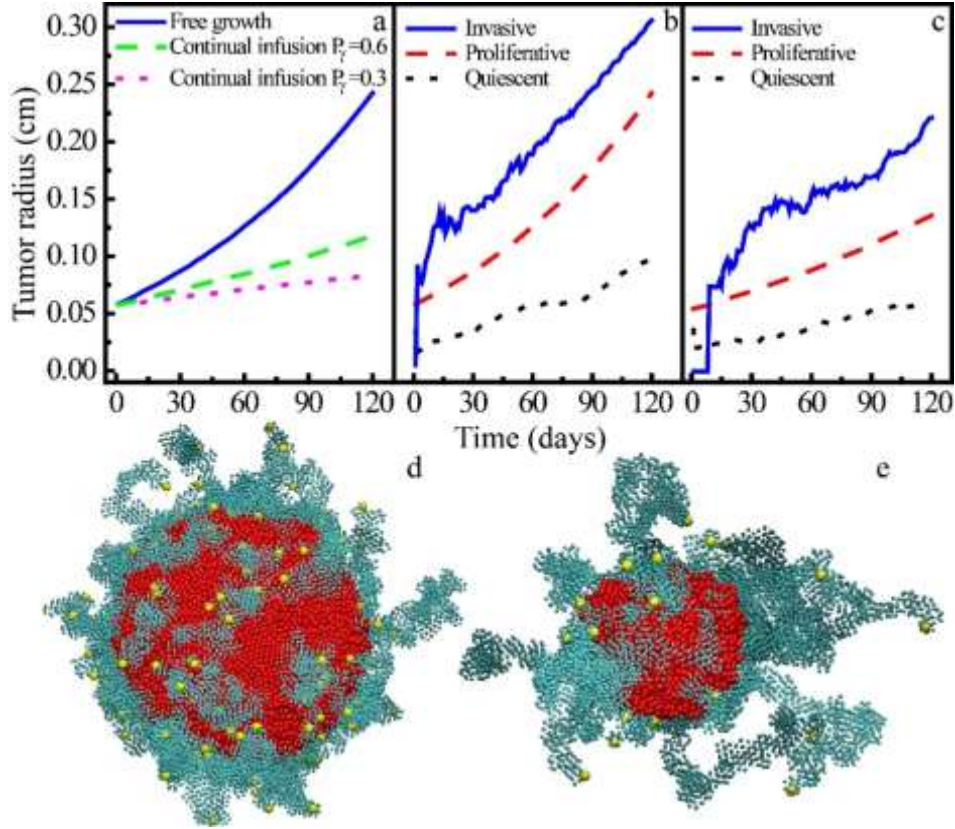

**Fig. 4 Tumor size and morphology under constant dosing condition.** (a) The growth dynamics of proliferative tumors with different drug concentrations in homogeneous microenvironment. (b) The average sizes of invasive tumor cells as a function of growth time without chemotherapy. (c) The average sizes of invasive tumor cells as a function of growth time with a division reduction factor $P_\gamma^0 = 0.6$. (d) A snapshot of the simulated growing tumor without chemotherapy on day 120 (with 142 invasive cells); (e) A snapshot of the simulated growing tumor under constant dosing condition ($P_\gamma^0 = 0.6$) on day 120 (with 31 invasive cells). As stated in the context, in this figure the micro-environment can be either avascular or vascularized. In (d) and (e), only the proliferative cells (red), the invasive cells (yellow) and the degraded ECM cells (cyan) are plotted.

Next, we consider invasive tumor growth under the constant dosing condition with a high drug concentration ($P_\gamma^0 = 0.6$). The mutation rate and mobility of the invasive cells are respectively set as 0.05 and 3 following Ref. [28]. The ECM degradation ability is 0.4. Fig. 4(b) and (c) shows the average linear size associated with the quiescent region, proliferative rim and invasive branches as growth time for a freely growing invasive tumor and one under chemotherapy for purposes of comparison. Snapshots of the morphology of the growing invasive tumors are also shown in Fig. 4(d) and (e).

We can see from Fig. 4(b) and (c) that when the drug is applied, the expansion of both the proliferative tumor and the invasive cells are apparently surprised. A more quantitative comparison shows that although proliferative cells grow much slower (the final primary tumor size is decreased by 50%) with drug infusion, the growth of the invasive cells is only weakly suppressed (the final extent size is decreased by 20%). In the chemo treated tumors, the invasive cells can still develop long invasive branches (see Fig. 4(e)) compared to the free growth case. The apparent decrease of the overall extent of the invasive branches is due to the significantly reduced size of the primary tumor. In fact the average linear extent of the invasive cells remains roughly the same as in the free growth case. This is because the drug does not affect the motility and ECM degradation of the invasive cells. However, the number of invasive cells is reduced by applying the drug, which is again due to the reduced division rate of the proliferative cells (i.e., less mutant daughter cells with invasive phenotype are produced).

**3.2.2 Invasive tumor growth under periodic dosing condition**

We now investigate the effects of drugs on the tumor growth under periodic dosing condition. In this condition, drugs are periodically applied and the drug concentration decays in a different manner in vascularized and avascular micro-environments as predicted by the pharmacokinetic calculations. Specifically, for the vascularized micro-environment, the drug concentration at the simulation domain boundary drops very quickly due to the fast transport via blood vessels; and for the avascular micro-environment, the drug concentration at the tumor boundary decays relatively slowly due to diffusion.

Fig. 5(a) shows the growth dynamics of both proliferative and invasive tumors in vascularized micro-environment under periodic dosing conditions. The distribution of drug concentration within the tumor during one proliferative cycle is discussed in Sec. 3.1.2. Different dosages are applied, which leads to different cellular division reduction factors i.e., $P_\gamma$ = 0.6; 0.3 and 0.05 used in the simulations. For purpose of comparison, we also consider the growth proliferative tumor without chemotherapy and with two different constant dosing conditions ($P_\gamma$ = 0.6 and 0.3).

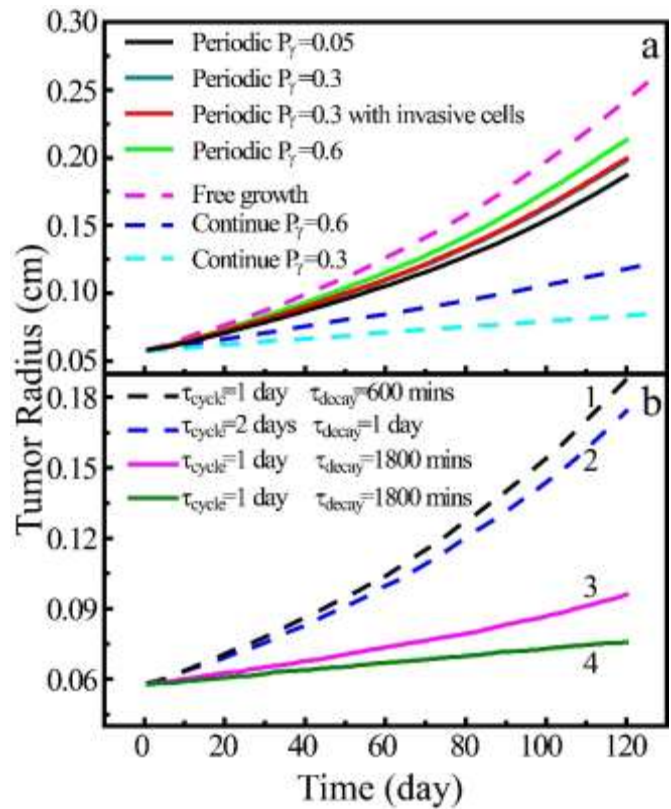

**Fig. 5: Rumor size in homogeneous microenvironment under periodic dosing conditions.** (a) Growth dynamics of the tumor in vascularized homogeneous microenvironment under different periodic dosing conditions. Effects of different dosages are modeled via different division reduction factor $P_\gamma$, with a dosing period and decay time of $\tau_{decay}$ = 1 day and $\tau_{decay}$ =600 min, respectively. For purpose of comparison, the results for constant dosing with $P_\gamma$ = 0.6 and 0.3 as well as freely growing tumors are also shown. (b) Growth dynamics of the invasive tumor in avascular homogeneous microenvironment under different periodic dosing conditions. The dosing period $\tau_{cycle}$, decay time $\tau_{decay}$ are shown in the figure. Here a division reduction factor $P_\gamma$ = 0.05 is used. The results for tumor in vascularized micro-environment with small decay time (cases 1 and 2) are also shown for purpose of comparison with the tumors in avascular micro-environment (cases 3 and 4). The consumption parameters used are $K_{met} = 2.0 \times 10^{-4}$ min$^{-1}$, $\lambda_{14} = 2.5 \times 10^{-4}$ s$^{-1}$ (cases 1, 2, and 3) and $2.5 \times 10^{-5}$ s$^{-1}$ (case 4).

We can see from Fig. 5(a) that in general periodic dosing conditions do not lead to the strong suppression of tumor growth as in the constant dosing cases, even for the high drug concentration cases (with a division reduction factor $P_\gamma = 0.05$). This is because for the periodic dosing, the drug concentration drops quickly in the vascularized micro-environment, which results in a weaker reduction of cellular division. This is to contrast with the constant dosing condition, which has been shown to be able to significantly suppress tumor growth in both vascularized and avascular micro-environments. However, such dosing condition can cause also damages to the normal cells and serious side effects. Thus, an alternative strategy is to infuse the drug more frequently with a shorter dosing period, as we will show below.

The growth dynamics of an invasive tumor under periodic dosing condition with $P_\gamma = 0.3$ is also shown in Fig. 5(a). We see that the growth curve of the primary tumor almost coincides with that of the proliferative tumor with the same $P_\gamma$. This is because although the invasive cells leave the primary tumor and migration into surrounding tissues, the division rate of the proliferative cells in both cases are almost identical, leading to the same primary tumor sizes.

We now investigate the effects of different periodic dosing conditions on the growth of invasive tumors in avascular micro-environment. Fig. 5(b) shows the growth dynamics of the primary tumor for different dosing period $\tau_{cycle}$ and decay time $\tau_{decay}$ but the same division reduction factor $P_\gamma = 0.05$. As shown in Fig. 5(b), cases 1 and 2 are associated the small decay time ($\tau_{cycle} > \tau_{decay}$), corresponding to the tumors in vascularized micro-environment for which the drug concentration decays very quickly due to fast pharmacokinetics. Cases 3 and 4 are associated with the large decay time, corresponding to the tumors in avascular micro-environment. It can be clearly seen that in the latter cases (i.e., avascular cases) the drug can effectively suppress the tumor growth. The reason is that for the avascular micro-environment the drug decay is much slower (also see Fig. 1(c)). This results in a higher drug concentration to suppress the tumor cell division. In addition, a large consumption parameter ($\lambda_{14} = 2.5 \times 10^{-4}$ s$^{-1}$) is used for case 3, and a small consumption parameter ($\lambda_{14} = 2.5 \times 10^{-5}$ s$^{-1}$) is used for case 4. We see that with a smaller consumption parameter, the drug concentration can remain a higher for a longer time, which leads to continuously suppression of tumor cell division and thus, a smaller final tumor size.

Finally we investigate the invasive tumor morphology under these periodic dosing conditions. Fig. 6(a) to (c) shows the snapshots of the growing tumors at day 120 under periodic dosing with different decay times, i.e., a fast decay with $\tau_{decay}$ = 600 min for (a) corresponding to tumors in vascularized micro-environment; and a slow decay with $\tau_{decay}$ = 1800 min for (b) and (c) corresponding to tumors in avascular micro-environment. We see that in all cases, there are a large number of dendritic invasive branches composed of collectively migrating invasive cells. Since the drug only reduces the proliferative cells' division rate, the linear extents of the invasive branches in all cases are almost the same. The sizes of the tumors in avascular micro-environment in Fig. 6(b) and (c) are smaller than the tumors in vascularized micro-environment in Fig. 6(a), which is due to the higher effective drug concentration in the avascular systems. The small consumption in (c) also leads to a decreased size than that in (b).

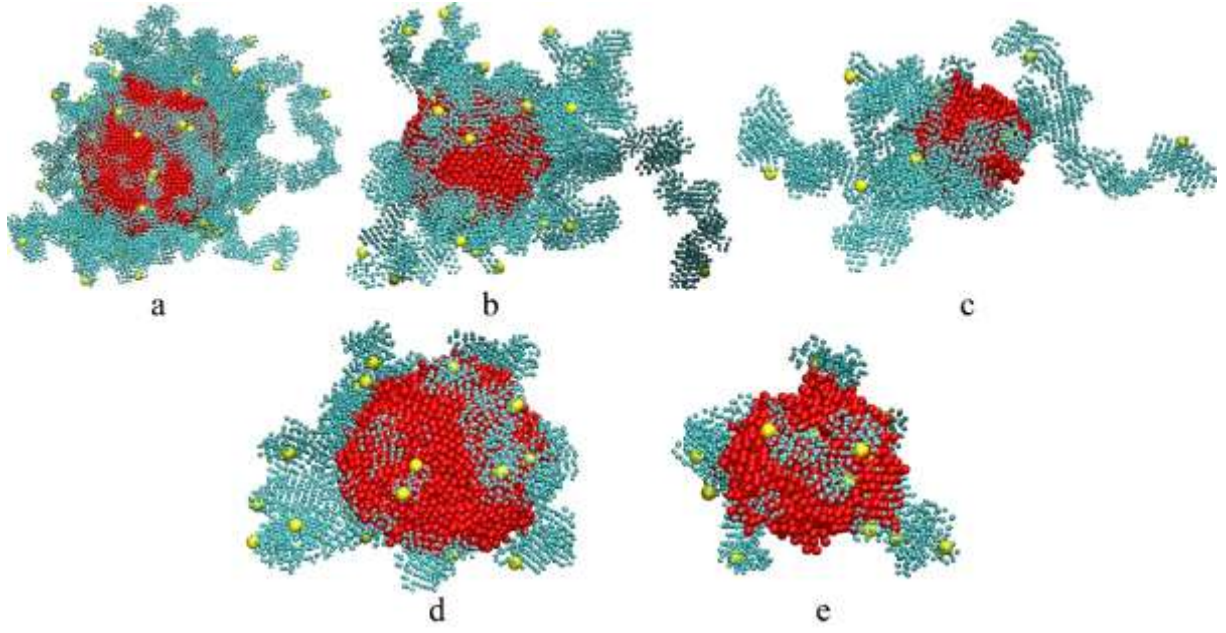

**Fig. 6 Snapshots of simulated tumors.** In all these cases, the same dosing period $\tau_{cycle}$ = 1 day, chemical decomposition parameter $K_{met}$ = 2.0 × 10$^{-4}$ min$^{-1}$, and division reduction factor $P_\gamma$ = 0.05 are used. Here only the proliferative cells (red), the invasive cells (yellow) and the degraded ECM cells (cyan) are plotted. (a)-(c): Snapshots of the simulated invasive tumors in homogeneous microenvironment under periodic dosing conditions at day 120. (a) is the tumor in vascularized micro-environment with a quick drug decay ($\tau_{decay}$ = 600 min) due to fast pharmacokinetics and a small consumption parameter (2.5 × 10$^{-5}$ s$^{-1}$); (b) is the tumor in avascular micro-environment with a slow drug decay ($\tau_{decay}$ = 1800 min) and a large consumption parameter (2.5 × 10$^{-4}$ s$^{-1}$); (c) is the tumor in avascular micro-environment with a slow drug decay ($\tau_{decay}$ = 1800 min) and a small consumption parameter (2.5 × 10$^{-5}$ s$^{-1}$). (d)-(e): Snapshots of the simulated tumors in heterogeneous microenvironment (with a random distribution of ECM density). (d) The ECM density ranges from 0.1 to 0.5 and with an average value of 0.3; (e) The ECM density ranges from 0.1 to 0.9 and with an average value of 0.5.

### 3.3 Evolution of invasive tumor in heterogeneous environment under periodic dosing conditions

In this section we employ our hybrid model to investigate invasive tumor growth in heterogeneous environment under periodic dosing conditions. To accurately capture the diffusion dynamics of drugs in the heterogeneous stroma, we explicitly utilize the location-dependent diffusion coefficient in the diffusion-reaction equation, i.e.,

$$\frac{\partial \phi(\vec{r},t)}{\partial t} = \nabla \cdot \left[ D(\vec{r},t) \nabla \phi(\vec{r},t) \right] - K_{met} \phi(\vec{r},t) - \lambda_0 \frac{\phi(\vec{r},t)}{\phi(\vec{r},t) + \phi_0} \quad . \tag{11}$$

The discretized form of the heterogeneous diffusion term in Eq. (11) is given below

$$\nabla \cdot \left[ D(\vec{r},t) \nabla \phi(\vec{r},t) \right] \Rightarrow$$

$$\frac{D_{i,j,k}}{2} \Delta(\phi^n_{i,j,k}) + \frac{D_{i+1,j,k}}{2}(\phi^n_{i+1,j,k} - \phi^n_{i,j,k}) + \frac{D_{i-1,j,k}}{2}(\phi^n_{i-1,j,k} - \phi^n_{i,j,k}) +$$

$$\frac{D_{i,j+1,k}}{2}(\phi^n_{i,j+1,k} - \phi^n_{i,j,k}) + \frac{D_{i,j-1,k}}{2}(\phi^n_{i,j-1,k} - \phi^n_{i,j,k}) + \frac{D_{i,j,k+1}}{2}(\phi^n_{i,j,k+1} - \phi^n_{i,j,k}) + \frac{D_{i,j,k-1}}{2}(\phi^n_{i,j,k-1} - \phi^n_{i,j,k})$$
(12)

The diffusion coefficient in a heterogeneous ECM depends on the local ECM density, which also represents the rigidity of the system in our model. Following the formulation of heterogeneous gas diffusion coefficient in systems with non-uniform pressures[73], we use the following empirical expression for $D(\vec{r},t)$

$$D(\vec{r},t) = \begin{cases} D_0 \cdot \rho_{0ECM}/(\rho_{ECM}(\vec{r},t) + \rho_{res}) & (\vec{r} \text{ lies in the ECM region}) \\ \eta D_0 & (\vec{r} \text{ lies in the tumor region}) \end{cases}$$
(13)

where $D_0$ is the diffusion coefficient in the uniform ECM possessing a density of $\rho_{0ECM}$, $\rho_{res}$ is the residual density after the ECM is completely degraded by the invasive cells. In the following simulations, we employ a random distribution for the ECM density and choose $\rho_{0ECM} = 0.3$, which is the value used for the homogeneous ECM in previous sections. And we choose $\rho_{res} = 0.1$. In addition, we set $\eta = 0.75$ since the diffusion in the tumor region is evidently slower due to high cellular density than that in the ECM.

In the heterogeneous environment, we use the following sinusoidal-like distribution to describe the initial (time=0) ECM density

$$\rho_{ECM}(\vec{r},0) = \rho_{ECM,av} - \frac{1}{2}\rho_{ECM,fluc} + \rho_{ECM,fluc} |\sin(\frac{\pi x}{L_x} + \frac{\pi y}{L_y} + \frac{\pi z}{L_z})|$$
(14)

where $\rho_{ECM,av}$ is the average ECM density, $\rho_{ECM,fluc}$ is the ECM fluctuation amplitude, $L_x(L_y, L_z)$ is the sinusoidal period. In our simulation, we set $\rho_{ECM,av} = 0.3$ for Fig. 6(d) and $\rho_{ECM,av} = 0.5$ for Fig. 6(e). $L_x(L_y, L_z)$ is about $L/60$ or $L/30$, where $L$ is the size of the cubic simulation box.

| $P_\gamma$ | $\lambda_{14}$ (s$^{-1}$) | $\rho_{ECM}$ | $\tau_{cycle}$ (day) | $\tau_{decay}$ (min) | $R_P$ *1 (cm) | $R_{Inv}$ *2 (cm) | $N_{Inv}$ *3 |
|---|---|---|---|---|---|---|---|
| 0.6 | 2.5*10$^{-4}$ | 0.3 | 1 | 600 | 0.210 | ------- | -------- |
| 0.3 | 2.5*10$^{-4}$ | 0.3 | 1 | 600 | 0.200 | ------- | -------- |
| 0.05 | 2.5*10$^{-4}$ | 0.3 | 1 | 600 | 0.198 | ------- | -------- |
| 0.05 | 2.5*10$^{-4}$ | 0.3 | 1 | 600 | 0.187 | 0.264 | 92 |
| 0.05 | 2.5*10$^{-4}$ | 0.3 | 2 | 1440 | 0.174 | 0.262 | 101 |
| 0.05 | 2.5*10$^{-5}$ | 0.3 | 1 | 600 | 0.158 | 0.236 | 72 (a) |
| 0.05 | 2.5*10$^{-4}$ | 0.3 | 1 | 1800 | 0.106 | 0.210 | 25 (b) |
| 0.05 | 2.5*10$^{-5}$ | 0.3 | 1 | 1800 | 0.076 | 0.182 | 9 (c) |
| 0.05 | 2.5*10$^{-5}$ | [0.1,0.5] | 1 | 1800 | 0.119 | 0.165 | 28 (d) |
| 0.05 | 2.5*10$^{-5}$ | [0.1,0.9] | 1 | 1800 | 0.091 | 0.118 | 16 (e) |

*1: The averaged radius of proliferative cells at day 120;

*2: The averaged radius of invasive cells at day 120;

*3: The number of invasive cells at day 120.

**Table 2: Summary of the model parameters for different periodic dosing conditions (the division decay factor $P_\gamma$; the consumption parameter $\lambda_{14}$; the two periodic dosage times $\tau_{cycle}$ and $\tau_{decay}$) as well as the growth data of tumors in both vascularized and avascular micro-environment (the averaged radius of proliferative cells and invasive cells $R_p$ and $R_{Inv}$; the number of invasive cells $N_{Inv}$) on day 120 in the heterogeneous ECM. The brackets in the last column indicate the corresponding morphology plots in Fig. 6.**

Snapshots of the morphology of both proliferative and invasive tumors growing in heterogeneous ECM with random density under periodic dosing are shown in Fig. 6(d) and (e). It can be clearly seen that in the heterogeneous ECM, the tumors develop rough and bumpy surface due to position dependent inhomogeneous cell division probability, as well as varying division reduction factors due to inhomogeneous drug concentration. This phenomenon has also been observed in the previous work for tumors growing in heterogeneous ECM with high rigidity [28].

Table 2 provides a detailed summary of the model parameters for different periodic dosing conditions as well as the growth data of tumors in both vascularized and avascular micro-environments on day 120 in the heterogeneous ECM. We can clearly see that for periodic dosing, the treatment is more effective in suppressing tumors in avascular heterogeneous micro-environment than that in vascularized heterogeneous ECM, consistent with the cases in homogeneous ECM. In addition, we find that denser and more rigid ECM (e.g., with average density 0.5) leads to an overall smaller tumor. However, under the same dosing condition, the tumor growing in heterogeneous ECM becomes more malignant (with larger primary tumor size and more invasive cells) compared the tumor growing in homogeneous ECM with the same density ($\rho_{ECM}$ = 0.3).

We note that a high ECM density, on the one hand, can suppress tumor growth; and on the other hand, can slow down the drug diffusion to the tumor region, which promotes tumor growth. Therefore, the actual tumor growth dynamics in heterogeneous ECM is the outcome of these two competing effects. For the case of average $\rho_{ECM}$ = 0.3, the diffusion of the drug is significantly slowed down while the density is not high enough to sufficiently suppress tumor growth, leading to a larger tumor compared with that in homogeneous ECM with the same density. For the case of an average $\rho_{ECM}$ = 0.5, the ECM density is large enough to suppress tumor growth even with very little drugs, and thus, results in a smaller tumor compared to that growing in corresponding homogeneous ECM. However, the invasiveness of the tumor growing in heterogeneous ECM with high density is significantly enhanced, which is consistent with the observation reported in Ref. [28].

### 3.4 Tumor growth dynamics in heterogeneous environments and non-uniform dosing conditions

To further demonstrate the utility and predictive capability of our hybrid model, we now examine the effects of periodic dosing on the growth dynamics of proliferative tumors in highly heterogeneous microenvironment. Specially, we consider two distinct cases for the environmental heterogeneities: (i) geometrically confined microenvironment and (ii) spatially non-uniform drug dosing.

### 3.4.1 Effects of geometrically confined microenvironment

Certain tumors such as ductal carcinoma in situ (DCIS) grow in a geometrically confined microenvironment, which usually results in a highly anisotropic tumor shape. On the other hand, the heterogeneity of the microenvironment also significantly influences the diffusion of drugs to the tumor site. Here, we apply our hybrid model to investigate the effects of periodic dosing on proliferative tumors growing in confined environment. The effects of the environmental confinement are modeled by a discontinuous distribution of ECM density. In particular, we consider the simulation domain is composed of two equal-sized sub regions. The left sub region possesses a higher ECM density value ($\rho_{ECM}$ = 0.6, e.g., to mimic the hard basal membrane) and the right region possesses a lower ECM density value ($\rho_{ECM}$ = 0.2, to mimic soft tissue). The dosing period is $\tau_{cycle}$ = 1 day, and the drug decay constant is $\tau_{decay}$ = 800 mins. The drug is released at the boundary of the spherical simulation domain with radius $R$ = 0.3 cm (see Fig. 1), which imposes a uniform initial high drug concentration at (and outside) the simulation boundary and zero concentration within the simulation domain. An initial tumor of linear size 0.06 cm is introduced in the center of the domain. The drug diffusion coefficient, which is a function of ECM density and local cell density, is obtained using Eq. (13).

The spatial-temporal evolution of the drug concentration in the ECM-tumor system is obtained by numerically solving Eq. (11). Figure 7(a) shows the drug concentration distribution in the *x-y* plane associated with $z$ = 0 at $t$ = 1.0 hour. Due to the high ECM density (i.e., low drug diffusivity) in the left region of the simulation domain, a higher concentration is built up compared to that in the right region. This leads to an overall non-symmetric distribution of drug in the system. However, due to the small tumor cell population (and size) at this stage (shown as the red circle in Fig. 7(a)), the difference in drug concentration in the left and right region next to the tumor is relatively small. Figure 7(b) shows the distribution of the division reduction factor $P_{r,\varphi}$ in the proliferative rim at $t$ = 15 days. We note that the observed fluctuations in $P_{r,\varphi}$ is mainly due to the heterogeneous cell division time in our hybrid model. As discussed in Sec. 2.4, in our CA model we consider a proliferative cell can divide at any time during a dosing cycle, implying a random distribution of cell division time. Under periodic dosing condition, the drug concentration at a cell at the time of division is generally different from that of another cell, leading to the non-uniform distribution of $P_{r,\varphi}$.

Figure 7(c) and (d) show the snapshots of the growing tumors (at $t$ = 150 days) in the confined environment with two different cell division probabilities ($p_0$ = 0.192 and 0.288). It can be clearly seen that the tumor develops a highly anisotropic shape, indicating the majority of proliferation occurs in the right low-density ECM region. On the other hand, protrusion-like structures are developed across the soft-hard ECM boundary, which is a key feature of microenvironment-enhanced malignancy. We note that even after 150 days, the protrusions remain relative compact. This is to contrast the elongated dendritic protrusions typically found in tumors growing in drug-free hard ECM [20, 21]. These observations illustrate the effects of drug treatment on proliferative tumor in confined microenvironment.

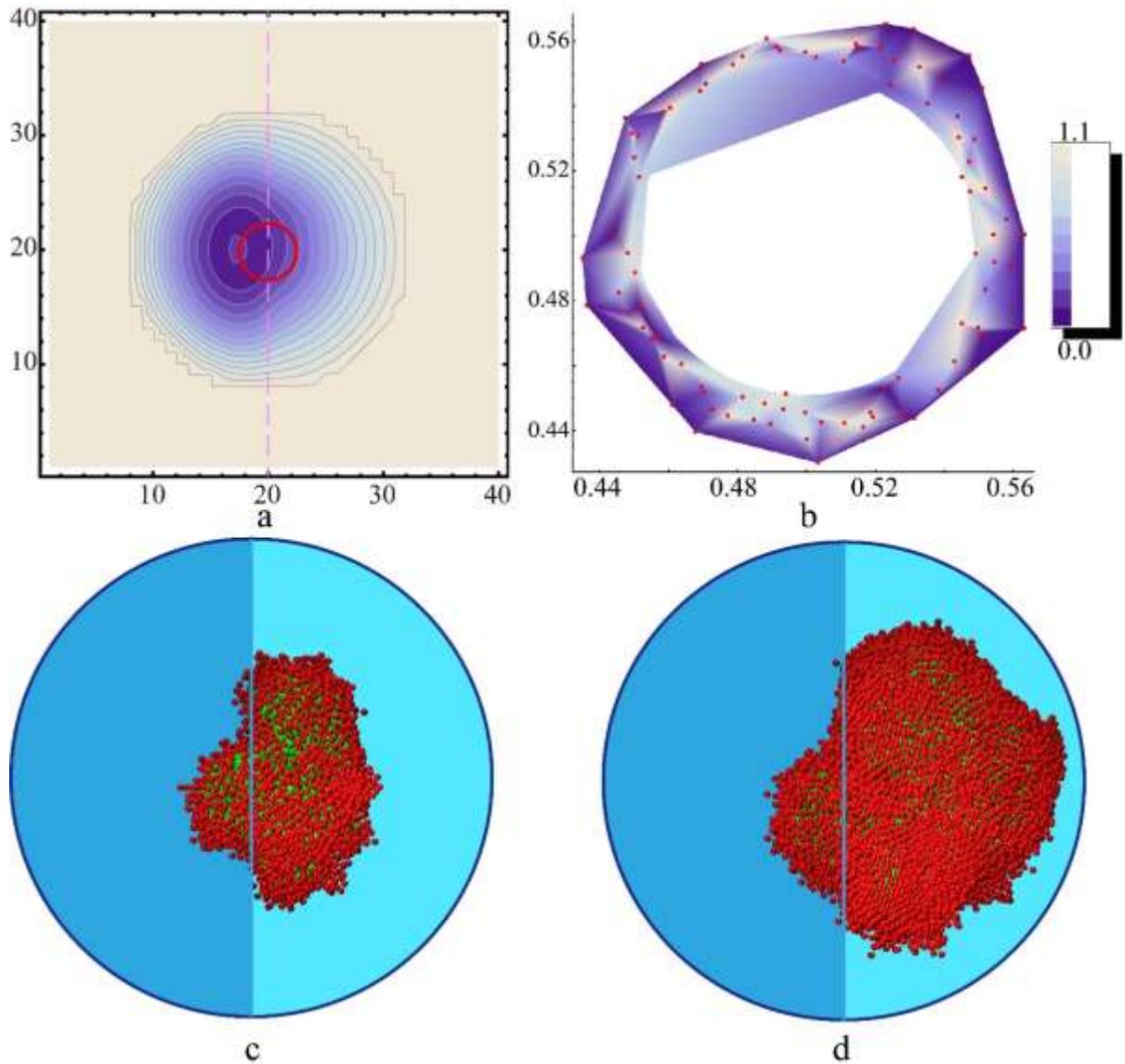

**Fig. 7 Effects of geometrically confined microenvironment on proliferative tumor growth under periodic dosing ($\tau_{decay}$ = 600 min, $\tau_{cycke}$ = 1 day).** (a) Asymmetric distribution of drug concentration in the *x-y* plane associated with *z* = 0 at *t* = 1.0 hour. the small red circle denotes the original tumor; (b) Distribution of the division reduction factor $P_{r,\varphi}$ in proliferative cells in the *x-y* plane associated with z = 0 on day 15. The red dots denote the proliferative cells. (c) Snapshot of a proliferative tumor growing in the confined microenvironment with $p_0$ = 0.192 on day 150; (d) Snapshot of a proliferative tumor growing in the confined microenvironment with $p_0$ = 0.288 on day 150. In (a), (c), (d), the middle lines denote the interfaces between two different ECM densities ($\rho_{ECM}$ is 0.6 in the left side and 0.2 in the right side)

### 3.4.2 Effects of spatially non-uniform drug dosing

Finally, we consider the effects of spatially non-uniform drug dosing. This is motivated by the fact that at certain stage of development, tumor cells can produce vascular endothelial growth factor (VEGF) to recruit endothelial cells for angiogenesis. The newly formed blood vessels can transport both nutrients and drugs to the local tumor site close to the blood vessels. When sufficient amount of drugs are transported to the tumor site, its local growth can be suppressed. Based on these considerations, in our simulation, instead of considering uniformly distributed vascular network (or avascular drug diffusion) on the tumor boundary as shown in Fig. 1, we consider the blood vessels recruited by the growing tumor are located in the lower left region of the tumor-ECM system. This is implemented by releasing the periodically dosed drugs at the lower left of the spherical simulation domain. Here, we set the drug concentration on the lower left region as a source boundary condition, where $\phi(\vec{r}, t)$ is the periodical function in Eq. (10) for only a restricted region, which satisfies the following conditions: (0.2 cm $\leq r \leq$ 0.3 cm; $x - x_0 <$ 0; $y - y_0 <$ 0), where $x_0 = y_0 = z_0 =$ 0.5 cm, are the coordinates of the spherical center.

Figure 8(a) shows the distribution of drug concentration in the *x-y* plane associated with $z =$ 0. It can be seen from Fig. 8(a) that after initial releasing, the drugs quickly diffuse to the entire system and are consumed and degraded. The lower left region of the domain remains to possess a relatively high drug concentration even after 16.8 hours of dosing, suggesting a high suppression of tumor growth in this region. Figure 8(b) shows the distribution of the division probability $P_{div}$ of the proliferative cells with $z =$ 0. It can be seen that the cells in the lower left region possess a much smaller division probability due to the high drug concentration in this region. Figure 8(c) and (d) show the snapshots of two proliferative tumors with different cell division probabilities ($p_0 =$ 0.192 and 0.384) at day 150. It can be seen that the effects of drugs are more significant in the fast growing tumor (Fig. 8(d)). Specifically, the cell division in the lower left region of the fast growing tumor is strongly suppressed by the high drug concentration, which results in a relatively flat edge in this region. On the other hand, the slowly growing tumor develops a relatively isotropic shape with a slightly flat edge in the lower left region.

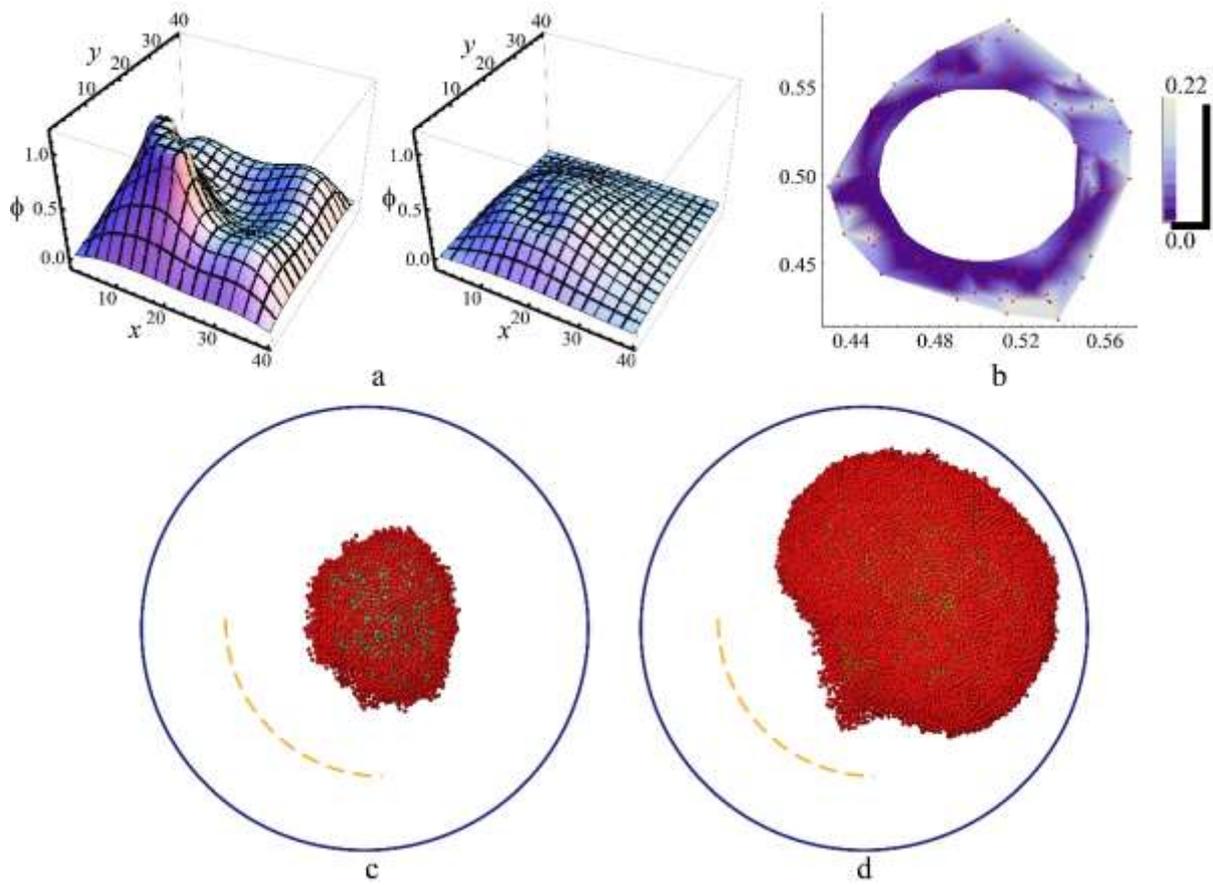

**Fig. 8 Effects of spatially non-uniform drug dosing ($\tau_{decay}$ = 600 min, $\tau_{cycke}$ = 1 day) on proliferative tumor growth ($\rho_{ECM}$ is 0.2 in the whole region).** (a) Evolution of drug concentration distribution in the $x$-$y$ plane associated with $z = 0$. The drugs are dosed in the lower left region of the spherical simulation domain periodically. Left panel: $t = 2.4$ hours; right panel: $t = 16.8$ hours. (b) Distribution of the division probability $P_{div}$ in proliferative cells in the $x$-$y$ plane associated with $z = 0$ on day 15. The red dots denote the proliferative cells. (c) Snapshot of a slowly growing tumor with $p_0 = 0.192$ on day 150. (d) Snapshot of a fast growing tumor with $p_0 = 0.384$ on day 150.

## 4. Summary and Conclusions

In this paper, we presented a comprehensive investigation of the effects of different chemotherapy

(i.e., constant vs. periodic dosing) on the growth dynamics of invasive tumors in both vascularized and avascular 3D heterogeneous microenvironment using a novel hybrid computational model. Our hybrid model integrates the physiologically based pharmacokinetic model for predicting overall drug concentration decay in different types of tumors, the continuum diffusion-reaction model for spatial-temporal evolution of the drug distribution in tumor-ECM system, as well as the discrete cell automaton model for invasive tumor growth simulation under effects of drugs. This model allows us to explicitly consider the effects of heterogeneous environment on drug diffusion, tumor growth and invasion as well as the drug-tumor interaction on individual cell level.

We have employed the hybrid model to investigate the evolution and growth dynamics of avascular invasive solid tumors in both vascularized and avascular micro-environments under chemotherapy with both constant and periodic dosing. We find that constant dosing is generally more effective in suppressing primary tumor growth compared to periodic dosing, due to the resulting continuous high drug concentration. Moreover, periodic dosing is found to be more effective in suppressing tumor growth in avascular micro-environment, due to the slower pharmacokinetics in such systems. However, as the chemotherapy is assumed not to suppress invasive cell migration, complex invasion branches emitting from the primary tumor have been found. In addition, we find that the malignancy of the tumor is significantly enhanced in highly heterogeneous microenvironment, leading to inefficient chemotherapy. We also use this model to the geometry-confined environment and non-uniform drug dosing situation. Our computational model, once supplemented with sufficient clinical data, could eventually lead to the development of efficient in silico tools for prognosis and treatment strategy optimization.

In our current model, the drug-tumor interaction is modeled as a reduction of the division probability (rate) of individual tumor cells, which depends on the local drug concentration. We note that this treatment does not explicitly consider the heterogeneity in the distribution of cell division time and cycle time and has assumed uniform distributions for these quantities. In future, we will further generalize our hybrid model to explicitly take into account the aforementioned heterogeneity, which would lead to a more accurate prediction of tumor growth dynamics under periodic dosing conditions.

## Acknowledgements

This work was supported by the State Key Development Program for Basic Research of China (Grant No 2013CB837200), the National Natural Science Foundation of China (Grant No 11474345 and 11674043, and11604030), and Arizona State University start-up funds.

## Author contribution

HX, YJ, YY, JS, RL and LL conceived and designed the model. HX and YJ performed the simulations. HX, QF and RL analyzed the data. HX, YJ, QF, MH, ZH, JY, GC, RL and LL wrote the paper.

## Additional Information

The authors declare no competing financial interests.